\documentclass[11pt,twoside]{article} 
\usepackage{asp2004}
\usepackage{psfig} 
\usepackage{epsf} 
\usepackage{graphics}
\usepackage{lscape} 
\def\edcomment#1{\iffalse\marginpar{\raggedright\sl#1\/}\else\relax\fi}
\reversemarginpar

\begin{document}
\title{Changes in the structure of the accretion disc of  V2051~Ophiuchi through the outburst cycle}
\author{R. F. Santos$^1$, R. Baptista$^2$, M. Faundez-Abans$^3$}
\affil{$^1$$^2$Universidade Federal de Santa Catarina, Brazil\\
$^3$LNA, Brazil}

\begin{abstract}

We present the results of the analysis of light curves of V2051 Oph
through an outburst with eclipse mapping techniques.
\end{abstract}
\thispagestyle{plain}

\section{Introduction} 

\hskip 0.55truecmDwarf novae show recurrent outbursts (of 2-5 mags on
timescales from weeks to months) powered by a sudden increase in mass
inflow in the accretion disc around the white dwarf.  Currently, there
are two competing models to explain the cause of the sudden increase
in mass accretion. In the mass transfer instability model (MTIM), the
outburst is the time dependent response of a viscous accretion disc to
a burst of matter transferred from the secondary star. In the disc
instability model (DIM), matter is transferred at a constante rate to
a low viscosity disc and accumulates in an annulus until a critical
configuration switches the disc to a high viscosity regime and the gas
diffuses rapidly inwards and onto the white dwarf. Tracking the evolution 
of a dwarf novae accretion disc along an outburst cycle with eclipse mapping
techniques (EMT) provides a valuable opportunity to test
these models againt observations.   

\section{Observations and data analysis}

V2051 Oph is a short period dwarf nova ($P_{orb}=90\,min$) with deep
eclipses ($B\simeq 2.5\,mag$). It went in outburst between the nights
of 2000 July 30 and 31.

B band light curves of V2051 Oph were obtained with the 1.60\,m
telescope of Laboratorio Nacional de Astrofisica (LNA/Brazil) on 2000
July 28/August 02 covering the nights before the onset of the
outburst, the short ($\approx1$\,day) outburst maximum and 2 days along the
decline from maximum.

The disc radius shrinks from $0.47\,R_{L1}$ in quiescence to
$0.40\,R_{L1}$ at the night before maximum, in agreement with the
expectation of the MTIM ($R_{L1}$ is the distance from the disc centre
to the inner Lagrangian point).

\section{Results}

\hskip 0.55truecm EMT were 
used to solve for a map of the disc
brightness distribuition and for the flux of an addional uneclipsed
component (Baptista\,\&\,Steiner\,1993).

The sequence of eclipse maps reveal that the outburst starts with a 
decrease in the brightness of the
disc and confirms that the disc shrinks at the night before
maximum. The map of this night is dominated by emission along the 
gas stream 
with no evidence of an increase in the brightness of the bright spot.
The brightness of the inner disc regions remains constant during
outburst maximum and along decline, while the brightness of the outer
disc regions progressively decreases with the inward propagation of a
cooling wave. The disc becomes fainter through the
decline phase, leaving the bright spot progressively 
more perceptible at the outer edge of the disc.

The maximum fractional contribution of the uneclipsed component (9\,\% of
the total flux in the B band) occurs on the night before outburst maximum.
This suggests the
development of a vertically-extended disc wind \emph{before} outburst maximum.

We quantify the changes in disc size during outburst by analyzing 
the radial intensity distribution of the total map. We define the outer
disc radius in each map as the radial position at which the intensity 
distribution falls below a re\-ference level corresponding to the
intensity of the bright spot in the eclipse map in
quiescence. The disc expands from $0.40\,R_{L1}$ at the night before maximum to
$0.76\,R_{L1}$ at outburst maximum. Two days after maximum, the disc
radius reduces to $0.67\,R_{L1}$. The radial intensity
distribution of the symmetric disc component shows an outward moving 
heating wave at the rise to 
outburst ma\-ximum with a speed of $v_{f(hot)}$$\,\geq\,1.73\,km\,s^{-1}$. 
The speed of the inward mo\-ving cooling wave are
\,$-0.24\,km\,s^{-1}$ and \,$-0.91\,km\,s^{-1}$, respectively 1 and 2 days 
after maximum. We observe an \emph{acceleration} of the
cooling wave as it travels across the disc, in contradiction
with the prediction of the DIM. From $v_{f(hot)}$ we derive a 
viscosity parameter $\alpha_{hot}\simeq$\,0.14\,, 
comparable to the derived viscosity in quiescence
$\alpha_{cool}\simeq$\,0.16 (Baptista\,\&\,Bortoletto\,2004).

The radial brightness temperature distribution is flatter than the $T
\propto R^{-3/4}$ law expected for steady-state discs both in
quiescence and in outburst, leading to larger mass accretion rates in
the outer disc\,($0.3-0.4\,R_{L1}$) than in the inner disc
regions\,($0.1\,R_{L1}$). If we assume a distance of 146\,pc to the
binary (Vrielmann,\,Stiening\,\&\,Offutt\,2002), the brightness
temperatures in quiescence range from 5800\,K in the outer disc to
9200\,K in the inner disc; at outburst maximum the temperatures range
from 9600\,K at $0.4\,R_{L1}$ to 12200\,K in the inner disc.  The
inferred temperatures of the outbursting disc are higher than the
critical temperature $T_{crit}$ above which the disc gas should remain
while in the high viscosity branch of the thermal-viscous limit cycle
of the DIM. However, if the distance is 100\,pc\,(see
Baptista\,\&\,Bortoletto\,2004), the inferred disc temperature are
lower and do not exceed $T_{crit}$ even at outburst maximum.

Our next step is to derive the distance to the binary from the measured B 
and V white dwarf fluxes
in quiescence.

\end{document}